\documentclass[11pt,twoside]{article}
\usepackage{psfig,amsmath}
\newcommand{\VolumeHeader}{}
\newcommand{\VolumeSerial}{LNS}

\newcommand{\ActivityName}{ {\normalsize {\it 
ICTP Summer School on Astroparticle Physics and Cosmology
}}}
\newcommand{\ActivityDate}{ {\normalsize {\it
Trieste, June 17-July 5, 2002
}}}

\newcommand{\be}{\begin{equation}}
\newcommand{\ee}{\end{equation}}
\newcommand{\bea}{\begin{eqnarray}}
\newcommand{\eea}{\end{eqnarray}}
\newcommand{\myequation}[1]{\begin{equation} {#1}  \end{equation} }
\newcommand{\gdm}{D}
\newcommand{\hdm}{{m\nu}}
\newcommand{\vis}{v}

\newcommand{\ApJ}{Astrophys. J}

\newcommand{\PRD}{Phys. Rev. D}

\newcommand{\aut}[2]{{#2.\ #1,}}
\newcommand{\laut}[2]{{#2.\ #1,}}
\newcommand{\refs}[6]{``#1,''  #2, {\bf #3},  {#4} (#5).}

\newcommand{\mybib}[2]{\bibitem{#2}}
\newcommand{\Spin}[4]{\, {}_{#2}^{\vphantom{#4}} {#1}_{#3}^{#4}}
\newcommand{\Gm}[3]{\, {}_{#1}^{\vphantom{#3}} G_{#2}^{#3}}

\newcommand{\LectureHeader}{Perturbation Theory}

\begin{document}
\pagestyle{myheadings}
\markboth{\LectureHeader}{\VolumeHeader}
\markright{\VolumeHeader}

\newcommand{\whit}[1]{#1}
\newcommand{\yell}[1]{#1}
\newcommand{\mage}[1]{#1}
\newcommand{\oran}[1]{#1}
\newcommand{\eqnsize}{}
\newcommand{\mytitle}[1]{{\bf #1}}
\newcommand{\mybullet}[1]{#1}
\newcommand{\mynbullet}[1]{#1}
\newcommand{\vertsp}{}
\newcommand{\potential}{{A}}
\newcommand{\shift}{{B}}
\newcommand{\curvature}{{H}_L}
\newcommand{\shear}{{H}_T}
\newcommand{\Spy}[3]{\, {}_{#1}^{\vphantom{#3}} Y_{#2}^{#3}}
\newcommand{\cmb}{\Theta}
\newcommand{\TT}{{\Theta\Theta}}
\newcommand{\len}{\phi}
\newcommand{\deld}{\delta}
\newcommand{\Rflat}{R}
\newcommand{\PP}{{\phi\phi}}
\newcommand{\EE}{{EE}}
\newcommand{\BB}{{BB}}
\newcommand{\TE}{{\Theta E}}

\newcommand{\like}{{\cal L}}
\newcommand{\bn}{\hat{\bf n}}
\newcommand{\bx}{{\bf x}}
\newcommand{\bk}{{\bf k}}
\newcommand{\bl}{{\bf l}}
\newcommand{\sig}[1]{{\boldsymbol \sigma}_{#1}}
\newcommand{\mat}[4]{\left( \begin{array}{cc} 
	#1 \! & #2 \\
	#3 \! & #4 \end{array} \right)}
	\newcommand{\intl}[1]{\int {d^2 {\bf l}_#1 \over (2\pi)^2}}
	\newcommand{\la}{{{\bf l}_1}}
	\newcommand{\lb}{{{\bf l}_2}}

\begin{titlepage}


\title{Covariant Linear Perturbation Formalism}

\author{Wayne Hu\thanks{whu@background.uchicago.edu}
\\[1cm]
{\normalsize
{\it CfCP, Department of Astronomy 
and Astrophysics, University of Chicago}}
\\[10cm]
{\normalsize {\it Lecture given at the: }}
\\
\ActivityName 
\\
\ActivityDate 
\\[1cm]
{\small \VolumeSerial} 
}
\date{}
\maketitle
\thispagestyle{empty}
\end{titlepage}

\baselineskip=14pt
\newpage
\thispagestyle{empty}


\begin{abstract}
Lecture notes on covariant linear perturbation theory and its applications to
inflation, dark energy or matter and the cosmic microwave background.
\end{abstract}

\vspace{6cm}



\newpage
\thispagestyle{empty}
\tableofcontents
\newpage
\section{Apologia}

In these informal {\it Lecture Notes} on formalism, 
I gather together a few elements of covariant linear perturbation
theory and its applications to inflation, dark energy or matter and the cosmic microwave background (CMB).
    I make no attempt to cite properly the original source
material.  My usual, lame defense is ``it's all in Bardeen's paper \cite{Bar80}'' --  go read it
and ignore these lecture notes!    

Seriously though, in addition to \cite{Bar80}, 
I have heavily relied on a few 
sources in compiling the various sections and refer the reader to references therein.
The covariant formalism presented in \S \ref{sec:formalism} and gauge transformation material
in \S \ref{sec:gauge}
draws from \cite{KodSas84} distilled in \cite{HuEis99}.  Applications to inflation in  \S \ref{sec:inflation}
draw from \cite{MukFelBra92}, to dark energy and matter in  \S \ref{sec:dark} from \cite{Hu98}, and to the CMB in  \S  \ref{sec:cmb} from \cite{HuSelWhiZal98}.  
For the last topic, despite great recent progress on the phenomenology, I have limited myself to the formal aspects that relate to covariant perturbation theory. 
My defense here is  ``everything that doesn't go out of date as soon as it's written is in Peebles \& Yu \cite{PeeYu70}'' -- go read it and stop looking for
CMB reviews!   My other defense is Matias Zaldarriaga was supposed to cover that in this {\it School} so blame him!  As for the equation density and opaqueness of these {\it Lecture Notes},  I have
no excuse -- take it as an homage to the Dick Bond lectures of my own student days -- at least
there are no figures with 100 curves.  As Martin White would say, you are physicists; you 
don't need figures!

\section{Formalism}
\label{sec:formalism}
\subsection{Covariant Approach}

Perturbation theory proceeds by linearizing the Einstein equations
\begin{eqnarray}
	\yell{G_{\mu \nu}} &=& 8\pi G \yell{T_{\mu \nu}}\,,
\end{eqnarray}
around a background metric.  
Here $G_{\mu\nu}$ is the Einstein tensor and $T_{\mu\nu}$ is the stress energy tensor.
The Bianchi identity $\nabla_{\mu} G^{\mu\nu}=0$ guarantees the covariant
conservation of the total stress-energy
\eqnsize\begin{eqnarray}
	\nabla_\mu \yell{T^{\mu\nu}} &=& 0\,.
\end{eqnarray}
The conservation equation is redundant with the Einstein equation but 
is a particularly useful representation of the equations when there are multiple
components to the matter that are separately conserved.

Let us begin by distinguishing between {\it covariant} equations and {\it invariant} variables:
\smallskip

\hskip 1.5cm \yell{Covariant} = equations takes same \yell{form} in all coordinate systems

\hskip 1.5cm \yell{Invariant} = variable takes the same \yell{value} in all coordinate systems
\smallskip

\noindent
In an evolving universe, the meaning of a density perturbation necessarily requires
the specification of the time slicing in relation to the background 
and hence there is no such thing as a gauge or coordinate
invariant density perturbation. Likewise for elements of the metric.     The general 
covariance of the Einstein and conservation equations,
on the other hand, can be preserved.  With covariant equations one can, after the fact of
their derivation, choose
the gauge that best suits a given physical problem, e.g. the evolution of inflationary, dark energy, 
or CMB fluctuations.

\subsection{Metric Representation}

Let us take the background metric to be the general homogeneous and isotropic, or
Friedmann-Robertson-Walker form.  Here the degrees of freedom are the
(comoving) spatial curvature $K$ and an overall scale factor for the expansion $a$
such that the line element
\begin{align}
ds^{2} &= g_{\mu\nu} dx^{\mu}dx^{\nu}= a^{2} ( -d\eta^{2} + \gamma_{ij}dx^{i} dx^{j} )\,.
\end{align}
The time variable is 
$\eta$ the conformal time and 
we  normalize the scale factor to $a=1$ today, $\eta(a=1)=\eta_{0}$.  
The three metric $\gamma_{ij}$ can be 
represented in spherical coordinates as 
\begin{align}
\gamma_{ij}dx^{i}dx^{j} & = d D^{2} + D_{A}^{2}d\Omega\,,
\end{align}
where $D$ is the comoving distance and 
$D_{A}=K^{-1/2} \sin( K^{1/2} D )$ is the angular diameter distance. 

A general perturbation to the FRW metric may be represented as 
\begin{align}
\yell{g^{00}} &= -a^{-2}(1-2 \yell{\potential})\,, \nonumber\\
\yell{g^{0i}} &= -a^{-2} \yell{\shift^i}\,,  \nonumber\\
\yell{g^{ij}} &= a^{-2} (\gamma^{ij} -
        2 \yell{\curvature} \gamma^{ij} - 2 \yell{\shear^{ij}})\,,
\label{eqn:metric}
\end{align}
yielding 
\mage{(1)} $\yell{A} \equiv$ a scalar \yell{potential}; 
\mage{(3)} $\yell{B^i}$ 
		a vector \yell{shift};
\mage{(1)} $\yell{H_L}$ a scalar perturbation to the spatial
		\yell{curvature}; 
\mage{(5)} $\yell{H_T^{ij}}$ a 
		\yell{trace-free}
		distortion to spatial metric, for a total of \mage{(10)} degrees of freedom.  This gives
		 a complete representation of the symmetric $4 \times 4$ metric tensor.

\subsection{Matter Representation}

\mybullet{Likewise expand the matter \yell{stress energy} tensor around a homogeneous density $\rho$ and pressure $p$ by introducing 10 degrees of freedom}
\eqnsize\begin{align}
\yell{T^0_{\hphantom{0}0}} &= -\rho - \yell{ \delta\rho}\,, \nonumber\vertsp\\
\yell{T^0_{\hphantom{0}i}} &= (\rho + p)(\yell{v_i} - \yell{\shift_i}) \,, \nonumber\vertsp\\
\yell{T_0^{\hphantom{i}i}} &= -(\rho + p)\yell{v^i}\,, \nonumber\vertsp\\
\yell{T^i_{\hphantom{i}j}} &= (p + \yell{\delta p}) \delta^i_{\hphantom{i}j} 
	+ p\yell{\Pi^i_{\hphantom{i}j}}\,, \vertsp
\label{eqn:stressenergy}
\end{align}
yielding
\mybullet{(1) \yell{$\delta\rho$} a scalar 
\yell{density perturbation}; (3) \yell{$v_i$} a vector \yell{velocity}, 
		(1) \yell{$\delta p$} a scalar \yell{pressure perturbation}; (5) \yell{$\Pi_{ij}$} a tensor \yell{anisotropic
			stress} perturbation}.
\mybullet{So far the treatment of the matter and metric
 is \yell{fully general} and applies to any type of matter or coordinate choice 
including non-linearities in the matter, e.g. cosmological defects.}

\subsection{Closure}

The Einstein equations do not specify a closed system of equations and require
the addition of supplemental conditions provided by the microphysics to fix
the dynamical degrees of freedom.  In the simplest
cases this closure is established through equations of state 
for the background and perturbations.
 
In counting the dynamical degrees of freedom, one must recall that the
conservation equations are redundant with the Bianchi identities and that
4 degrees of freedom are absorbed by the choice of coordinate system 
\eqnsize\begin{eqnarray}
\yell{20}  && \hbox{Variables (10 metric; 10 matter)} \nonumber\\
\yell{-10} && \hbox{Einstein equations} \nonumber\\
\yell{-4}   && \hbox{Conservation equations} \nonumber\\
\yell{+4}  && \hbox{Bianchi identities} \nonumber\\
\yell{-4}   && \hbox{Gauge (coordinate choice 1 time, 3 space}) \nonumber\\
\hbox{------}  
 && \nonumber\\
\yell{6} && \hbox{Degrees of freedom}\nonumber
\end{eqnarray}
\mybullet{Without loss of generality these six dynamical degrees of freedom 
can be taken to be the \yell{6 components} of the \yell{matter stress tensor}.}
		
Since the background is isotropic, the unperturbed matter is described entirely
by the pressure $p(a)$ or equivalently \yell{$w(a) \equiv p(a)/\rho(a)$}, the
\yell{equation of state} parameter.  For the perturbations specification of the relationship 
between $\delta p$, $\Pi$ and the other perturbations, e.g. $\delta \rho$ and $v$
suffices.

\subsection{Friedmann Equations}

The unperturbed Einstein equation yields the ordinary Friedmann equation
which relates the expansion rate to the energy density
\eqnsize\begin{align}
\left( \frac{\dot a}{a} \right)^2 + K & =   \frac{8\pi G}{3}a^{2} \yell{\rho} \,, 
\label{eqn:friedmann}
\end{align}
where overdots represent conformal time derivatives, and the acceleration
Friedmann equation which relates the change in the expansion rate to the equation
of state
\begin{align}
{d \over d \eta}\left( {\dot a \over a} \right)  &= -\frac{4\pi G}{3}a^{2}(\yell{\rho} + 3 \yell{p}) \,,
\label{eqn:acceleration}
\end{align}
\mynbullet{so that $\yell{w} \equiv p/\rho < \mage{-1/3}$ implies acceleration
of the expansion.}

The conservation law  $\nabla^\mu T_{\mu\nu} = 0$ implies
\eqnsize\begin{equation}
\frac{ \yell{\dot\rho} } { \yell{\rho} }  = -3(1+\yell{w}) \frac{\dot a}{a}\,,
\label{eqn:rhodot}
\end{equation}
which  by virtue of the Bianchi identity can be derived from the Friedmann equation.

The counting exercise for the background becomes
\eqnsize\begin{eqnarray}
\yell{20} && \hbox{Variables (10 metric; 10 matter)} \nonumber\\
\yell{-17} && \hbox{Homogeneity and Isotropy}\nonumber\\
\yell{-2} && \hbox{Einstein equations} \nonumber\\
\yell{-1}   && \hbox{Conservation equations} \nonumber\\
\yell{+1}  && \hbox{Bianchi identities} \nonumber\\
\hbox{------}  
 && \nonumber\\
\yell{1} && \hbox{Degree of freedom}\nonumber
\end{eqnarray}
\mybullet{Without loss of generality this degree of freedom can be chosen to 
 be the equation of state $\yell w(a) = p(a)/\rho(a)$.}

\subsection{Linearization and Eigenmodes}

For the inhomogenous universe,
the Einstein tensor $G_{\mu\nu}$ is in general constructed out of a nonlinear 
combination of metric fluctuations.   Metric fluctuations typically remain small 
even in the presence of larger matter fluctuations.  Hence we linearize the
left hand side of the Einstein equations to obtain a set of partial differential
equations that are linear in the variables.   

These equations may then be decoupled into a set of ordinary differential equations
by employing normal modes under translation and rotation.  
The scalar, vector and tensor eigenmodes of the 
\yell{Laplacian operator} form a complete set
\eqnsize\begin{equation}
\begin{array}{rcll}
\nabla^2 \yell{Q^{(0)}} &= &  -k^2 \yell{Q^{(0)}} \quad &{\rm \mage{S}} \,,\vertsp \\
\nabla^2 \displaystyle{\yell{Q_i^{(\pm 1)}}}
	   &= & -k^2 \yell{Q_i^{(\pm 1)}} \quad &{\rm \mage{V}}\vertsp \,, \\
\nabla^2 \yell{Q_{ij}^{(\pm 2)}}
	   &= &  -k^2 \displaystyle{\yell{Q_{ij}^{(\pm 2)}}} \quad \vertsp &{\rm \mage{T}} \,.  \\
\end{array}
\end{equation}
In a spatially flat ($K=0$) universe, the eigenmodes are essentially plane waves 
\eqnsize\begin{eqnarray}
\yell{ Q^{(0)} } &=& \yell{ \exp( i {\bf k} \cdot {\bf x}) } \,,  \nonumber\\
\yell{ Q_i^{(\pm 1)} }
	&=& \frac{-i}{\sqrt{2}} (\hat{\bf e}_1 \pm i \hat{\bf e}_2)_i 
	\yell{ \exp(i {\bf k}\cdot {\bf x})\,, }\nonumber\\
\yell{ Q_{ij}^{(\pm 2)} }
 	&=& - \sqrt{\frac{3}{8}}
(\hat{\bf e}_1 \pm i \hat{\bf e}_2)_i (\hat{\bf e}_1 \pm i \hat{\bf e}_2)_j
\yell{\exp(i {\bf k}\cdot {\bf x})}\,, 
 \end{eqnarray}
 where ${\bf e}_{1}, {\bf e}_{2}$ are unit vectors spanning the plane transverse
 to ${\bf k}$.   
  
\mybullet{Vector modes represent divergence-free vectors (vorticity); tensor modes
represent 
transverse traceless tensors (gravitational waves)}
\begin{align}
\nabla^i Q_i^{(\pm 1)} & = 0\,,\qquad
\nabla^i Q_{ij}^{(\pm 2)}  = 0\,,\qquad
\gamma^{ij} Q_{ij}^{(\pm 2)}  = 0\,. 
\end{align}
Curl free vectors and the longitudinal components of tensors are 
represented with covariant derivatives of the scalar and vector modes
\eqnsize\begin{align}
\yell{Q_i^{(0)}} & =  -k^{-1} \nabla_i \yell{Q^{(0)}}\,, \nonumber \\
\yell{Q_{ij}^{(0)}} & =   
	(k^{-2} \nabla_i \nabla_j + {1 \over 3} \gamma_{ij}) 
	\yell{Q^{(0)}}\,,  \nonumber\\
\yell{Q_{ij}^{(\pm 1)}} & =   -{1 \over 2k}[ \nabla_i \yell{ Q_j^{(\pm 1)} }
	+ \nabla_j \yell{Q_i^{(\pm 1)}}] \,,  
\end{align}

\mybullet{For the $k$th eigenmode, the \yell{scalar components}  become}
\eqnsize\begin{equation}
\begin{array}{rclrcl}
\yell{\potential ({\bf x})} &=& \yell{\potential(k)} \, Q^{(0)} 
	\,, \quad& \yell{\curvature ({\bf x})}  &=& \yell{\curvature(k)} \, Q^{(0)} \,, \\
\yell{\delta\rho ({\bf x})} &=& \yell{\delta\rho (k)}\, Q^{(0)}\,, \quad& \yell{\delta p({\bf x})}  
 &=& \yell{\delta p (k)} \, Q^{(0)} \,,
\end{array}
\end{equation}
\mynbullet{the \yell{vectors components} become}
\eqnsize\begin{equation}
\begin{array}{rclrcl}
\yell{\shift_i ({\bf x})} & =& \displaystyle{\sum_{m=-1}^1} \yell{ B^{(m)}(k) } \, Q_i^{(m)}\,, \quad &
\yell{v_i({\bf x})} & =& \displaystyle{\sum_{m=-1}^1} \yell{ v^{(m)}(k)} \, Q_i^{(m)}\,,
\end{array}
\end{equation}
\mynbullet{and the \yell{tensors components} become}
\eqnsize\begin{equation}
\yell{ \shear{}_{ij}({\bf x})}  =  \sum_{m=-2}^{2}  \yell{ H_T^{(m)}(k) }\, Q_{ij}^{(m)}  \,,\quad
\yell{ \Pi_{ij}({\bf x})}  =  \sum_{m=-2}^2 \yell{  \Pi^{(m)}(k)} \,  Q_{ij}^{(m)}\,.
\end{equation}
An arbitrary set of spatial perturbations can be formed through a superposition of the
eigenmodes given their completeness.
  
\subsection{Covariant Scalar Equations}

The \yell{Einstein equations} for the scalar modes (suppressing $0$ superscripts) 
become
\begin{align}
 (k^2 - 3K)[ \yell{H_L}+ {1 \over 3} \yell{H_T}+ {\dot a \over a}
	 ({\yell{B} \over k}-  { \dot H_T \over k^{2} } )]  & = 4\pi G a^2  \left[ \yell{\delta \rho} + 3 {\dot a \over a} (\rho+p){\yell{v}-
\yell{B} \over k}\right] \,,
\nonumber\\
  k^2 ( \yell{A} + \yell{H_L} + {1 \over 3} \yell{H_T} ) + \left({d \over d\eta}+ 2 {\dot a \over a} \right)
	(k \yell{B} -  \yell{\dot H_T}) 
& = -8\pi G a^2 p \yell{\Pi}  \,,
\nonumber\\
{\dot a \over a} \yell{A} - \yell{\dot H_L} 
-{1 \over 3}  \yell{\dot H_T} - {K \over k^2} (k \yell{B}- \yell{\dot H_T})
&=  4\pi G a^2 (\rho+p){\yell{v}-\yell{B} \over k} \,,
\nonumber\\
\left[2 {\ddot a \over a} - 2 \left( {\dot a \over a} \right)^2 + {\dot a \over a} {d \over d\eta}
- {k^2 \over 3}\right] {\yell A} 
- \left[ {d\over d\eta} + {\dot a \over a} \right] (\yell{\dot H_L} +  {k \yell{B} \over 3})
&= 4\pi G a^2 (\yell{\delta p} + {1 \over 3}\yell{\delta\rho} ) \,.
\label{eqn:poisson}
\end{align}
The conservation equations become the continuity and Navier-Stokes equations
\begin{align}
\left[{d \over d\eta} + 3 {\dot a \over a}\right] \yell{\delta\rho}
	+  3{\dot a \over a} \yell{\delta p}
&=
        -(\rho+p)(k \yell{v} + 3\yell{\dot H_L})\,,  \label{eqn:continuity}\\
\left[ {d \over d\eta} + 4{\dot a \over a}\right] (\rho + p){\yell{v}-\yell{B} \over k} 
&= 
  \yell{ \delta p }- {2 \over 3}(1-3{K\over k^2})p \yell{\Pi} + (\rho+ p) \yell{A} \,.
  \label{eqn:Euler}
\end{align}
In the absence of anisotropic stress, the Navier-Stokes equation is known as the 
Euler equation.

These equations are not independent due to the Bianchi identity.  For numerical
solutions it suffices to retain 2 Einstein equations and the 2 conservation laws.
The counting exercise becomes 
\eqnsize\begin{eqnarray}
\yell{8} && \hbox{Variables (4 metric; 4 matter)} \nonumber\\
\yell{-4} && \hbox{Einstein equations} \nonumber\\
\yell{-2}   && \hbox{Conservation equations} \nonumber\\
\yell{+2}  && \hbox{Bianchi identities} \nonumber\\
\yell{-2}   && \hbox{Gauge (coordinate choice 1 time, 1 space}) \nonumber\\
\hbox{------}  
 && \nonumber\\
\yell{2} && \hbox{Degrees of freedom}\nonumber
\end{eqnarray}
Without loss of generality we may choose the dynamical components to be
those of the \mage{stress
tensor\yell{ $\delta p$, $\Pi$}.  Microphysics defines their relationship to the
density and velocity perturbations.   In fluid dynamics this is the familiar need of a prescription
for viscosity and heat conduction (entropy generation) 
to close the system of equations. }

\subsection{Covariant Vector Equations}

The vector Einstein equations become 
\begin{align}
\label{eqn:Poissonvector}
 (1-{2K \over k^2} )(k\yell{B^{(\pm 1)}} - \yell{\dot H_T^{(\pm 1)}}) 
& = 16\pi G a^2 
(\rho+p){\yell{v^{(\pm 1)}}-\yell{B^{(\pm 1)}} \over k} \,,\nonumber\\
 \left[ {d \over d\eta} + 2 {\dot a \over a} \right] 
(k\yell{B^{(\pm 1)}} - \yell{\dot H_T^{(\pm 1)}}) 
&= -8\pi G a^2 p\yell{\Pi^{(\pm 1)}}\,,
\end{align}
and the conservation equations become
\begin{align} \left[{d \over d\eta}+4{\dot a \over a}\right]
(\rho + p){\yell{v}^{(\pm 1)}-\yell{B}^{(\pm 1)} \over k} 
& = - {1 \over 2}(1-2{K \over k^2})p\yell{ \Pi^{(\pm 1)}}\,.
\label{eqn:Eulervector}
\end{align}
Since gravity provides \yell{no source} to vorticity, any initial vector perturbation
will simply decay unless vector anisotropic stresses are continuously generated in the matter.
In the absence of nonlinearities in the matter, e.g.\ from cosmological defects, vector
perturbations can generally be ignored.

The counting exercise for vectors becomes
\eqnsize\begin{eqnarray}
\yell{8} && \hbox{Variables (4 metric; 4 matter)} \nonumber\\
\yell{-4} && \hbox{Einstein equations} \nonumber\\
\yell{-2}   && \hbox{Conservation equations} \nonumber\\
\yell{+2}  && \hbox{Bianchi identities} \nonumber\\
\yell{-2}   && \hbox{Gauge (coordinate choice 2 space)}) \nonumber\\
\hbox{------}  
 && \nonumber\\
\yell{2} && \hbox{Degrees of freedom}\nonumber
\end{eqnarray}
\mybullet{Without loss of generality, we can choose these to be the
vector components of the \mage{stress
tensor}  \yell{$\Pi^{(\pm 1)}$}.}

\subsection{Covariant Tensor Equations}

The Einstein equation for the tensor modes is
\begin{equation}
\left[ {d^2 \over d\eta^2} + 2 {\dot a \over a}{d \over d\eta} + (k^2+2K) \right]
\yell{H_T^{(\pm 2)}} = 8\pi G a^2 p \yell{\Pi^{(\pm 2)}} \,,
\label{eqn:Poissontensor}
\end{equation}
and the counting exercise becomes
\eqnsize\begin{eqnarray}
\yell{4} && \hbox{Variables (2 metric; 2 matter)} \nonumber\\
\yell{-2} && \hbox{Einstein equations} \nonumber\\
\yell{-0}   && \hbox{Conservation equations} \nonumber\\
\yell{+0}  && \hbox{Bianchi identities} \nonumber\\
\yell{-0}   && \hbox{Gauge (coordinate choice 1 time, 1 space}) \nonumber\\
\hbox{------}  
 && \nonumber\\
\yell{2} && \hbox{Degrees of freedom}\nonumber
\end{eqnarray}
\mybullet{without loss of generality we can choose these to be
represented by the tensor components of the \mage{stress
tensor} \yell{  $\Pi^{(\pm 2)}$.}}

In the absence of anisotropic stresses and spatial curvature, the tensor equation becomes
a simple source-free gravitational wave propagation equation
\begin{equation}
\yell{\ddot H_T^{(\pm 2)}}  +
2 {\dot a \over a}
\yell{\dot H_T^{(\pm 2)}}  +
k^2\yell{H_T^{(\pm 2)}} = 0 \,,
\label{eqn:PoissontensorFree}
\end{equation}
which has solutions
\begin{align}
H_T^{(\pm 2)}(k\eta) & = C_1 H_1(k\eta)  + C_2 H_2(k\eta)\,, \\
H_1(x) &\propto x^{-m} j_m(x) \nonumber \,, \\
H_2(x) &\propto x^{-m} n_m(x) \nonumber\,,
\end{align}
\mynbullet{where $m=(1-3w)/(1+3w)$.}
\mybullet{If $w>-1/3$ then the gravitational wave amplitude 
is constant above horizon $x \ll 1$ and then oscillates
	and damps.}
If $w<-1/3$ then gravity wave oscillates and freezes into some value.  It will be useful to recall these solutions when 
	considering the Klein Gordon equation for scalar field fluctuations
	during inflation and dark energy domination.

\subsection{Multicomponent universe}

With multiple matter components, the Einstein equations of course remain valid
but with the associations
\begin{align}
\delta \rho &= \sum_{J} \delta \rho_{J}\,,\nonumber\\
(\rho+p)v^{(m)} &= \sum_{J} (\rho_{J}+p_{J}) v_{J}^{(m)}\,, \nonumber\\
p \Pi^{(m)} & = \sum_{J} p_{J} \Pi_{J}^{(m)}\,,
\end{align} 
where  $J$ indexes the different components, e.g. photons, baryons,
neutrinos, dark matter, dark energy, inflaton, cosmological defects.   
The conservation equations remain
valid for each non-interacting subsystem.  Interactions can be represented
as non-ideality in the specification of the stress degrees of freedom
(e.g. viscous and entropic terms)  in 
each subsystem or by explicitly writing down energy and momentum exchange
terms in separate conservation equations (see e.g. \S \ref{sec:cmb}). 

\section{Gauge}
\label{sec:gauge}

\subsection{Semantics}

The covariant equations of the last section hold true under any coordinate
choice for the relationship between the unperturbed FRW background and the
perturbations.  Choice of a particular relationship, called a gauge choice can
help simplify the equations for the physical conditions at hand.   The price 
to pay is that the perturbation variables are quantities that take on the meaning
of say a density perturbation or a spatial curvature perturbation only on a specific
gauge -- but that is in any case unavoidable.   

Since the preferred gauge for
simplifying the physics can change as the universe evolves, it is often useful
to access the perturbation variables of one gauge from those of another gauge.
This operation proceeds by writing down a covariant form for gauge specific variables
through the properties of gauge transformation.  It is the analogue of deriving covariant
equations for the dynamics but for the perturbation variables themselves.
Since these gauge specific variables
are now thought of in a coordinate independent way, this procedure is called  in
the literature endowing a variable with a  ``gauge invariant'' meaning.  
Note neither the numerical values nor the physical interpretation of the
variables has changed.  ``Gauge invariance`` here
is an operational distinction that indicates a freedom to calculate gauge specific 
quantities from an arbitrary gauge. 
Perturbation variables still only take on their given meaning in  the given gauge.
Under this  definition of gauge invariance, the perturbation variables of any fully specified
gauge is gauge invariant.   We will hereafter avoid using this terminology. 

To reduce the opportunity for confusion, we will name the metric perturbations in 
the various common gauges separately.   We will however rely on context to distinguish
between matter variables on the various gauges.

\subsection{Gauge Transformation}

In an evolving inhomogeneous universe, metric and matter fluctuations take on
different values in different coordinate systems.  Consider
a general coordinate transformation
\eqnsize\begin{eqnarray}
 \eta &=&\tilde \eta + \yell{T}\,, \\
x^i   &=& \tilde x^i + \yell{L^i}\,. \nonumber
\end{eqnarray}
Under this general transformation, the metric and stress energy tensors transform
as tensors.  The elements of these tensors are the perturbation variables and
their numerical values change with the transformation.

The coordinate choice represented by \yell{ $(T,L^i)$ }  can be decomposed
into scalar and vector modes.  
With $L^{(0)}=L$, the scalar mode variables
transform as
\begin{align}
 A &=\tilde A - \yell{\dot T} - {\dot a \over a} \yell{T}\,, \nonumber\\
 B &=  \tilde B + \yell{\dot L} + k\yell{T} \,, \nonumber\\
 H_L &=  \tilde H_L - {k \over 3}\yell{L} - {\dot a \over a} \yell{T}\,, \nonumber\\
 H_T &=  \tilde H_T + k\yell{L}\,, 
\label{eqn:metrictrans}
\end{align}
for the metric and 
\begin{align} 
{\delta \rho_J} &= {\delta\tilde\rho}_J - \dot\rho_J \yell{T}, \nonumber\\ 
{\delta  p_J} &= {\delta \tilde p}_J -\dot p_J \yell{T}, \nonumber\\
 v_J &=  \tilde v_J + \yell{ \dot L}
\label{eqn:fluidtrans}
\end{align}
for the matter.

For the vector mode variables
\begin{align}
 B^{(\pm 1)} &= \tilde B^{(\pm 1)} + \yell{\dot L^{(\pm 1)}}\,,\nonumber\\
 H_T^{(\pm 1)} &= \tilde H_T^{(\pm 1)} + k\yell{L^{(\pm 1)}}\,,\nonumber\\
 v_J^{(\pm 1)} &=  \tilde v_J^{(\pm 1)} + \yell{ \dot L^{(\pm 1)}}\,. 
\end{align}
The tensor mode variables are invariant under the gauge transformation as are
the components of the anisotropic stress tensor. 

A gauge is \yell{fully specified} if there is an explicit prescription for
		\yell{$(T,L)$} for scalars and  $(L^{(\pm 1)})$ for the vectors to 
		get to the desired frame.  Gauges are typically defined by conditions
		on the metric or matter fluctuations.  We now consider several common 
		scalar gauge choices and their uses.		
		
\subsection{Newtonian Gauge}

The Newtonian or longitudinal gauge is defined by diagonal metric fluctuations 
\begin{align}
	 B &=  H_T=0\,, \nonumber\\
	\yell{\Psi} &\equiv  A  \quad \hbox{\mage{(Newtonian potential)}} \,,\nonumber\\
	\yell{\Phi} &\equiv  H_L \quad \hbox{\mage{(Newtonian curvature)}} \,,
\end{align}
where $\Psi$ plays the role of the gravitational potential in the Newtonian approximation
and $\Phi$ is the Newtonian spatial curvature.  
This condition completely fixes the gauge by giving explicit expressions for the
gauge transformation
\begin{align}	
	L      &= - {\tilde H_T \over k} \,,\nonumber\\
	T      &= -{\tilde B \over k} + {1\over k^{2}}{d \over d\eta}{\tilde  H_T}\,.
\end{align}
The Einstein equations become
\begin{align}
(k^2 -3K)\yell{\Phi} &= 4\pi G a^2 \left[ \yell{\delta \rho} + 3 \frac{\dot a}{a}(\rho+p){\yell{v} \over k}\right] \,,\nonumber\\
k^2(\yell{\Psi} + \yell{\Phi}) &= -8\pi G a^2 p \yell{ \Pi}\,.
\end{align}
Note that for scales inside the Hubble length $k (\dot a /a)^{-1} \gg 1$, the first equation becomes the
Poisson equation for the perturbations and if the anisotropic stress of the
matter is much less than the energy density perturbation (as in the case of
non-relativistic matter), $\Psi \approx -\Phi$.

The conservation laws for the $J$th non-interacting subsystem becomes
\begin{align}
\left[{d \over d\eta} + 3 {\dot a \over a}\right] \yell{\delta\rho_{J}}
	+  3{\dot a \over a} \yell{\delta p_{J}}
&=
        -(\rho_{J}+p_{J})(k \yell{v_{J}} + 3\yell{\dot\Phi})\,,  \label{eqn:Newtcontinuity}\\
\left[ {d \over d\eta} + 4{\dot a \over a}\right]  (\rho_{J} + p_{J}) {\yell{v_{J}} \over k}
&= 
  \yell{ \delta p_{J} }- {2 \over 3}(1-3{K\over k^2})p_{J}\yell{\Pi_{J}} + (\rho_{J}+ p_{J}) \yell{\Psi} \,.\label{eqn:NewtEuler}
\end{align}
Aside from the $\dot \Phi$ term, these are the usual relativistic fluid equations.  
Since $\Phi$ represents a perturbation to the scale factor, $\dot \Phi$ represents
the perturbation to the redshifting of the density in an expanding universe
[see Eqn.~(\ref{eqn:rhodot})].

The Newtonian gauge is useful since it most closely corresponds to Newtonian gravity.
A drawback is that as $k (\dot a /a)^{-1} \rightarrow 0$ there are relativistic corrections to
the Poisson equation which make a straightforward implementation of the equations
numerically unstable.  Likewise in this limit, the metric effects on density 
 though $\dot \Phi$ muddle the interpretation of the conservation law.

\subsection{Comoving Gauge}

The comoving gauge is defined by 
\begin{align}
	B &=  v \quad (T^0_i=0)\,,\nonumber\\
	H_T      &= 0\,, \nonumber\\
	\yell{\xi}          &=  A \,,\nonumber\\
	\yell{\zeta}	     &=  H_L \quad\mage{\hbox{(comoving curvature})}\,,  
\end{align}
which completely fixes the gauge through
\begin{align}
	T	     &= (\tilde v-\tilde B)/k\,, \nonumber\\
	L    	     &= -\tilde H_T/k \,.
\end{align}
The Einstein equations become
\begin{align}
\yell{\dot \zeta} + Kv/k -  {\dot a \over a}\xi & = 0\,, \nonumber\\
\dot v + 2 {\dot a \over a}  v + k (\zeta+\xi) &= -8\pi G a^{2} p \Pi\,,
\end{align}
and the conservation laws become
\begin{align}
\left[{d \over d\eta} + 3 {\dot a \over a}\right] \yell{\delta\rho_{J}}
	+  3{\dot a \over a} \yell{\delta p_{J}}
&=
        -(\rho_{J}+p_{J})(k \yell{v_{J}} + 3\yell{\dot\zeta})\,,  \label{eqn:comcontinuity}\\
\left[ {d \over d\eta} + 4{\dot a \over a}\right] (\rho_{J} + p_{J}) {\yell{v_{J}}-v \over k}
&= 
  \yell{ \delta p_{J} }- {2 \over 3}(1-3{K\over k^2})p_{J}\yell{\Pi_{J}} + (\rho_{J}+ p_{J}) \yell{\xi} \,.\label{eqn:comEuler}
\end{align}
In particular the Navier-Stokes equation for the total matter becomes an algebraic relation
between total stress fluctuations and the potential
\begin{equation}
{(\rho+p) \xi = - \delta p + {2 \over 3} \left( 1 - {3K \over k} \right) p \Pi}
\end{equation}
so that these equations are a complete set.

Eliminating $\xi$ allows us to write a conservation law for the  
comoving curvature
\begin{equation}
\yell{\dot \zeta} + Kv/k = {\dot a \over a} \left[ -\frac{\yell{\delta p}}{\rho+p} + \frac{2}{3}\left(1-\frac{3K}{k^2} \right)
\frac{p}{\rho+p}\yell{\Pi}\right]\,.
\label{eqn:zetadot}
\end{equation}
On scales well below the curvature scale, this equation states that the comoving
curvature only changes in response to stress gradients which move matter around.
This statement corresponds to the non-relativistic intuition that causality should prohibit
evolution above the horizon scale.    This conservation law is the fundamental virtue
of the comoving gauge.   An auxiliary consideration is that the comoving gauge
variables are numerically stable.

From an alternate gauge choice, one can construct the Newtonian variables via the 
gauge transformation relation
\begin{align}
\zeta = \tilde H_{L} + {1\over 3}\tilde H_{T} - {\dot a \over a}{\tilde v - \tilde B \over k}\,.
\label{eqn:zetacov}
\end{align}
For example, since in the Newtonian gauge $H_{T}$ and $v$ take on the
same values
\begin{align}
\Phi &= \zeta + {\dot a \over a} {v \over k} \nonumber\\
        &= \zeta + {2 \over 3}{1 \over 1+w}{1 \over 1+ K (\dot a /a)^{-2}} [\Psi - \dot \Phi/(\dot a /a)]\,.
\end{align}
If the curvature is negligible, stress perturbations are negligible, and the equation of state
$w$ is constant
\begin{align}
\Phi = { 3+ 3w \over 5+ 3w} \zeta\,,
\end{align}
so that $\Phi$ tends to be of the same order as $\zeta$ but changes when the equation
of state changes.   

The Newtonian curvature can also be obtained through the comoving density perturbations
via  
 \begin{equation}
(k^{2}-3K) \Phi   = 4\pi G a^{2}\delta \rho \Big|_{\rm comoving} \,,
\end{equation}
which has the added benefit of taking the form of a simple non-relativistic Poisson equation.
This Poisson equation allows us to rewrite the conservation equation for the comoving curvature in
the case of negligible background curvature and anisotropic stress as
\begin{align}
{d \ln \zeta \over d\ln a} &= -{\Phi \over \zeta} {2 \over 3 + 3w} k^{2} (\dot a /a)^{-2}
{\delta p \over \delta \rho} \Big|_{\rm comoving} \nonumber \\
& \approx  - {2 \over 5 + 3w} k^{2}(\dot a/a)^{-2} {\delta p\over \delta \rho} \Big|_{\rm comoving} \,.
\label{eqn:nonadiabatic}
\end{align} 
For adiabatic stresses where $\delta p = c_{a}^{2}\delta \rho$, the change in 
the comoving curvature is negligible for $c_{a}k(\dot a/a)^{-1} \sim
c_{a} k\eta  \ll 1$, i.e. on scales below the sound horizon.   

\subsection{Synchronous Gauge}

The synchronous gauge confines the metric perturbations to the spatial
degrees of freedom 
\begin{align}
	 A &=  B = 0\,, \nonumber \\
	\yell{\eta_T} &\equiv -\frac{1}{3}  H_T - H_{L}\,, \nonumber\\
  h_{L} &= 6H_L\,, \nonumber\\
	T &= a^{-1} \int d\eta a \tilde A + c_1 a^{-1} \nonumber\,, \\
	L &= - \int d\eta(\tilde B+kT) + c_2 \,.
\end{align}
The metric conditions do not fully specify the gauge and need to be supplemented
by a definition of $(c_{1},c_{2})$.   Usually one defines
$c_{1}$ through the condition that the dark matter has zero velocity in the initial
conditions and $c_{2}$ through the setting of the initial curvature perturbation.
With a completely specified gauge condition, the synchronous gauge is as valid
as any other gauge.  The variables $\eta_{T}$ and $h_{L}$ form a stable
system for numerical solutions and hence the synchronous gauge has been
extensively used in numerical solutions.

The Einstein equations give
\begin{align}
\dot\eta_{T} - {K \over 2 k^{2}}(\dot h_{L}+6\dot\eta_{T} ) &= 4\pi G a^{2}(\rho+p){v \over k} \,,\nonumber\\
\ddot h_{L} + {\dot a \over a}\dot h_{L} &= -8\pi G a^{2} (\delta \rho + 3\delta p) \,,
\end{align}
while the conservation equations give
\begin{align}
\left[{d \over d\eta} + 3 {\dot a \over a}\right] \yell{\delta\rho_{J}}
	+  3{\dot a \over a} \yell{\delta p_{J}}
&=
        -(\rho_{J}+p_{J})(k \yell{v_{J}} + {1 \over 2}\yell{\dot h}_{L})\,,  \label{eqn:synchcontinuity}\\
\left[ {d \over d\eta} + 4{\dot a \over a}\right] (\rho_{J} + p_{J}) {\yell{v_{J}} \over k}
&= 
  \yell{ \delta p_{J} }- {2 \over 3}(1-3{K\over k^2})p_{J}\yell{\Pi_{J}} \,.\label{eqn:synchEuler}
\end{align}
Note that the lack of a potential in synchronous gauge implies that there are no gravitational
forces in the Navier-Stokes equation.  Hence for stress free matter like cold dark matter, zero velocity
initially implies zero velocity always.

\subsection{Spatially Flat Gauge:}

Conversely the spatially flat gauge eliminates spatial metric perturbations
\begin{align}
	H_L &=  H_T = 0 \,, \nonumber\\
	\alpha_{F} & \equiv A \,,\nonumber\\
	\beta_{F} & \equiv B \,,\nonumber \\
	T & =  \left( \frac{\dot a}{a} \right)^{-1} \left( \tilde H_L + \frac{1}{3}\tilde H_T \right)\,.\nonumber\\
		L & =  -\tilde H_T/k \,, 
\end{align}
The Einstein equations give
\begin{align}
\dot \beta_{F} + 2{\dot a \over a}\beta_{F} + k\alpha_{F}& =-8\pi G a^{2} p\Pi/k \,,\nonumber\\
{\dot a \over a} \alpha_{F} - {K \over k}\beta_{F} &= 4\pi G a^{2} (\rho+p){v-\beta_{F} \over k}\,,
\end{align}
and the conservation equations give
\begin{align}
\left[{d \over d\eta} + 3 {\dot a \over a}\right] \yell{\delta\rho_{J}}
	+  3{\dot a \over a} \yell{\delta p_{J}}
&=
        -(\rho_{J}+p_{J})k \yell{v_{J}}\,,  \label{eqn:continuityflat}\\
\left[ {d \over d\eta} + 4{\dot a \over a}\right] (\rho_{J} + p_{J}) {\yell{v_{J}}-\beta_{F} \over k}
&= 
  \yell{ \delta p_{J} }- {2 \over 3}(1-3{K\over k^2})p_{J}\yell{\Pi_{J}} + (\rho_{J}+ p_{J})\yell{\alpha_{F}} \,.\label{eqn:eulerflat}
\end{align}
The spatially flat gauge is useful in that it is the complement of the comoving gauge.  
In particular the comoving curvature is constructed from Eqn.~(\ref{eqn:zetacov}) as
\begin{equation}
\zeta  = -{\dot a \over a} { v - \beta_{F}  \over k} \,.
\end{equation}
This gauge is most often used in inflationary calculations where $v-\beta_{F}$ is
closely related to perturbations in the inflaton field.

\subsection{Uniform Density Gauge:}

Finally, one can eliminate the density perturbation $\delta\rho=0$ with the choice
\begin{align}
 H_T & = 0 \,, \nonumber\\
\zeta_\delta & \equiv H_L \nonumber\\
B_\delta & \equiv B \nonumber\\
A_\delta & \equiv A \nonumber\\
T & = {\delta\tilde \rho\over \dot \rho} \nonumber\\
L & = -\tilde H_T/k
\end{align}
The Einstein equations become
\begin{align}
 (k^2 - 3K)[ \yell{\zeta_\delta}+  {\dot a \over a}
	 {\yell{B_\delta} \over k} ]  & = 12 \pi G a^2   {\dot a \over a} (\rho+p)
	  { \yell{v}-
\yell{B_\delta} \over k} \,,
\nonumber\\
{\dot a \over a} \yell{A_\delta} - \yell{\dot \zeta_c} 
- {K \over k} \yell{B_\delta}
&=  4\pi G a^2 (\rho+p){\yell{v}-\yell{B_\delta} \over k} \,,
\end{align}
from which $A_\delta$ may be eliminated in favor of $\zeta_\delta$ and
$B_\delta$.
The conservation equations become
\begin{align}
 \left[{d \over d\eta} + 3 {\dot a \over a}\right] \yell{\delta\rho_{J}}
	+  3{\dot a \over a} \yell{\delta p_{J}}&=
        -(\rho_J+p_J)(k \yell{v_J} + 3\yell{\dot \zeta_\delta})\,,  \nonumber\\
\left[ {d \over d\eta} + 4{\dot a \over a}\right] (\rho_J + p_J){\yell{v_J}-\yell{B_\delta} \over k} 
&= 
  \yell{ \delta p_J }- {2 \over 3}(1-3{K\over k^2})p_J \yell{\Pi_J} + (\rho_J+ p_J) \yell{A_\delta} \,.
\end{align}
Notice that the continuity equation for the net density perturbation becomes
a conservation  law
\begin{equation}
\dot \zeta_\delta = -{\dot a \over a} {\delta p \over \rho+p} - {1\over 3} kv \,.
\end{equation}
Furthermore since $\delta \rho=0$, $\delta p$ is the non-adiabatic stress
[see Eqn.~(\ref{eqn:nonadiabatic})] and the conservation law resembles that of
the comoving curvature [see Eqn.~(\ref{eqn:zetadot})].   
More specifically,
the two curvatures are related by
\begin{equation}
\zeta_\delta = \zeta +{1\over 3} {\delta\tilde \rho\over (\rho+p)} \Big|_{\rm comoving}\,.
\end{equation}
By the same argument as that following Eqn.~(\ref{eqn:nonadiabatic}), these
two curvatures coincide outside the horizon if the stresses are adiabatic.
Hence they are often used interchangeably in the literature.
It bears an even simpler relationship to density fluctuations in the
spatially flat gauge
\begin{equation}
\zeta_\delta ={1\over 3} {\delta\tilde \rho\over (\rho+p)}\Big|_{\rm flat}\,.
\end{equation}
For a single particle species $\delta \rho/(\rho+p) = \delta n/n$, the number density fluctuation.
These simple relationships make $\zeta_{\delta}$ and related variables useful for the consideration of
relative number density or isocurvature perturbations in a multicomponent system.

\section{Inflationary Perturbations}
\label{sec:inflation}

\subsection{Horizon,  Flatness,  Relics Redux}

Inflation was originally proposed to solve the horizon, flatness and relic problem: that
the cosmic microwave background (CMB) temperature is isotropic across scales larger
than Hubble length at recombination; that the spatial curvature scale is at least comparable
to the current Hubble length; that the energy density is not dominated by defect relics from 
phase transitions in the early universe.

Measurements of fluctuations in the CMB imply a stronger version of the 
problems.  They imply spatial curvature
fluctuations above the Hubble length and rule out  cosmological defects as the primary
source of structure in the universe, i.e.~they are absent as a significant component of
the
perturbations as well as the background. 
The  conservation of the comoving curvature above the comoving
Hubble length $(\dot a/a)^{-1} \ll 1$ leaves a paradox as to 
their origin.  As long as the Hubble length monotonically increases, there will always be
a time $\eta \sim k^{-1}$ before which $\zeta$ cannot have been generated
dynamically.  For sufficiently small $k$, this time exceeds the 
recombination epoch at $a \sim 10^{-3}$ where the CMB fluctuations are formed.  

The comoving Hubble length decreases if $\rho$ scales more slowly than
$a^{-2}$ or $w<-1/3$, i.e. if the expansion accelerates [see Eqn.~(\ref{eqn:friedmann}-\ref{eqn:acceleration})].   Vacuum energy or 
a cosmological constant can provide for acceleration but not for re-entry into
the matter-radiation dominated expansion.   On the other hand, a scalar field  has both a potential and kinetic energy and provides a mechanism
by which acceleration can end.  
A major success of the inflationary model is that the same mechanism that
solves the horizon, flatness and relic problems by making the observed universe today
much smaller than the Hubble length at the beginning of inflation {\it predicts} 
super-Hubble curvature perturbations from quantum fluctuations in the scalar field.
The evolution of scalar field fluctuations can be usefully described in the gauge covariant
framework.   
  
\subsection{Scalar Fields}

The covariant perturbation theory described in the previous sections also 
applies to the inflationary universe.  The stress-energy tensor of 
a scalar field rolling in the potential $V(\phi)$ is 
\begin{equation}
T^{\mu}_{\hphantom{\mu}\nu} = \nabla^\mu \varphi\, \nabla_\nu \varphi
        - {1 \over 2} (\nabla^\alpha \varphi\, \nabla_\alpha \varphi + 2 V)
        \delta^{\mu}_{\hphantom{\mu}\nu} \,.
\end{equation}
For the backround $\bar \phi  \equiv \phi_0$ and 
\begin{equation}
\rho_\phi = {1 \over 2} a^{-2} \dot\phi_0^2 + V\,, \quad
p_\phi =      {1 \over 2} a^{-2} \dot\phi_0^2 -V\,.
\end{equation}
If the scalar field is kinetic energy dominated
$w_\phi = p_\phi/\rho_\phi \rightarrow 1$,
whereas if it is potential energy dominated 
$w_\phi = p_\phi/\rho_\phi \rightarrow -1$.

The equation of motion for the scalar field is simply the energy conservation
equation
\begin{equation}
\dot\rho_\phi = - 3( \rho_\phi + p_\phi){\dot a \over a} \,,
\end{equation}
\mynbullet{re-written in terms of $\phi_{0}$ and $V$}
\begin{equation}
{\ddot \phi_0} + 2{\dot a \over a} \dot\phi_0 +a^2 V' = 0\,.
\end{equation}
\mybullet{Likewise for the perturbations $\phi = \phi_0 + \phi_1$}
\begin{align}
\delta \rho_\phi
&=a^{-2}(\dot\phi_0 \dot\phi_1-\dot\phi_0^2 A)
+V' \phi_1 \nonumber\,,\\
\delta p_\phi &=
a^{-2}(\dot\phi_0 \dot\phi_1-\dot\phi_0^2 A)
- V' \phi_1 \nonumber\,,\\
(\rho_\phi+p_\phi) (v_\phi-B) &= a^{-2} k \dot\phi_0 \phi_1
        \nonumber\,,\\
p_\phi \pi_\phi & =  0\,,
\label{eqn:scalarfieldpert}
\end{align}
\mybullet{the continuity equation implies}
        \begin{align}
\ddot \phi_1 &= - 2{\dot a \over a} \dot \phi_1
- (k^2 + a^2 V'')\phi_1  
+  (\dot A - 3\dot H_L - kB) \dot\phi_0
        - 2 A a^2 V' \,,
        \end{align}
        while the Euler equation expresses an identity. 

\subsection{Gauge Choice}

We are interested in inflation as a means of generating comoving curvature
perturbations and so naively it would seem that inflationary perturbations should
be calculated in the comoving gauge.   However for a scalar field dominated
universe, the comoving gauge sets $B=v=v_\phi$
and hence $\phi_{1}=0$.  In the comoving gauge, the scalar field carries no perturbations
by definition!  This fact will be useful for treating scalar fields as a dark energy candidate
in the next section: in the absence of scalar field fluctuations, energy density and pressure
perturbations come purely from the kinetic terms so that $\delta p_{\phi} = \delta \rho_{\phi}$
yielding stable perturbations within the Hubble length.

The gauge covariant approach allows us to compute the comoving curvature from
variables in another gauge.
Since a scalar field transforms as a scalar field
\myequation{\phi_1 = \tilde\phi_1 - \dot\phi_0 T}
the comoving gauge is obtained from an arbitrary gauge by the time
slicing change $T = \tilde \phi_{1} /\dot \phi_{0}$.  The comoving curvature becomes
\begin{align}
\zeta & = \tilde H_L - {k \over 3} L - {\dot a \over a}T \,,\nonumber\\
         & = \tilde H_L + {\tilde H_T \over 3} - {\dot a \over a} {\tilde \phi_1 \over \dot \phi_0}\,,
\end{align}
and hence the simplest gauge from which to calculate the comoving curvature is
one in which $H_{L}=H_{T}=0$, i.e. the spatially flat gauge.  In this case
\begin{equation}
\zeta = - {\dot a \over a}{\tilde \phi_1 \over \dot \phi_0}
\end{equation}
and so a calculation of $\tilde \phi_{1}$ trivially gives $\zeta$.
Notice that the proportionality constant involves $\dot \phi_{0}$.  The slower
the background field is rolling, the larger the curvature fluctuation implied by
a given field fluctuation.  The reason is that the time surfaces must be warped
by a correspondingly larger amount to compensate $\tilde \phi_{1}$.
We will hereafter drop the tildes and assume that the scalar field fluctuation
applies to the spatially flat gauge.

\subsection{Perturbation Evolution}

In the spatially flat gauge the scalar field equation of motion can be written in
a surprisingly compact form.  
Beginning with the spatially flat gauge equation
        \begin{align*}
\ddot \phi_1 &= - 2{\dot a \over a} \dot \phi_1
- (k^2 + a^2 V'')\phi_1  
+  (\dot \alpha_{F}  - k\beta_{F}) \dot\phi_0
        - 2 \alpha_{F} a^2 V' \,.
        \end{align*}
the metric terms may be eliminated through Einstein equations
\begin{align*}
 \alpha_{F} &= 4\pi G a^2 \left({\dot a \over a} \right)^{-1} (\rho_\phi + p_\phi) (v_\phi- \beta_{F}) /k \\
    & = 4\pi G \left( {\dot a \over a} \right)^{-1}  \dot \phi_0 \phi_1\,,
\end{align*}
\mybullet{and ($k^{2} \gg |K|$) }
\begin{align}
 k \beta_{F} & = 4\pi G a^2  \left( {\dot a \over a} \right)^{-1} \left[ \delta\rho_\phi + 3 {\dot a \over a} (\rho_\phi + p_\phi)(v_\phi - \beta_{F})/k \right]\nonumber\\
 & = 4\pi G \left[\left({\dot a \over a}\right)^{-1} ( \dot \phi_0 \dot \phi_1+ a^2 V' \phi_1) - \left({\dot a \over a}\right)^{-2} (4\pi G \dot\phi_0)^2 
\dot \phi_0 \phi_1 + 3 \dot \phi_0 \phi_1\right]
\end{align}
so that $\alpha_{F}$, $\dot \alpha_{F} - k\beta_{F} \propto \phi_1$ with proportionality that depends only on the
background evolution, i.e. the  Einstein and scalar field equations reduce to a single second order
differential equation with the form of an expansion damped oscillator
\begin{align}
\ddot \phi_1 + 2{\dot a \over a } \dot \phi_1 + [k^2 + f(\eta)]\phi_1 \,. 
\end{align}
The expression for $f(\eta)$ can be given explicitly in terms of 
the parameters
\begin{align}
\epsilon &\equiv {3 \over 2} ( 1 + w_\phi) = {\frac{3}{2}\dot \phi_0^2/ a^2 V  \over 1 + {1 \over 2} \dot \phi_0^2 / a^2 V}\,,
\end{align}
which represents the deviation from a de Sitter expansion and 
\begin{align}
\delta \equiv {\ddot \phi_0 \over \dot \phi} \left( {\dot a \over a} \right)^{-1} - 1\,,
\end{align}
which represents the deviation from the overdamped limit of $d^2 \phi_0 / dt^2 = 0$,
where $dt = a d\eta$.  
When small, these quantities are known as the slow-roll parameters.   
The Friedmann equations become
\begin{align}
\left({  \dot a \over a }\right)^2 &= 4\pi G \dot \phi_0^2 \epsilon^{-1}\,,\nonumber \\
{d \over d\eta}\left( {\dot a \over a} \right) &= \left({ \dot a \over a } \right)^2 (1-\epsilon)\,,
\end{align} 
\mybullet{and the background field equation becomes}
\begin{equation}
\dot \phi_0 {\dot a \over a} (3 + \delta ) = -a^2 V'\,,
\end{equation}
\mybullet{Together they yield the equation of motion for $\epsilon$}
\myequation{\dot \epsilon = 2 \epsilon ( \delta + \epsilon) {\dot a \over a}\,.}
It is straightforward now to show that defining  $u \equiv a \phi$ the perturbation equation 
becomes
\begin{align}
\ddot u + [k^2 +g(\eta)] u = 0\,, 
\label{eqn:inflationosc}
\end{align}
\mynbullet{where}
\begin{align}
g(\eta) & \equiv f(\eta) + \epsilon - 2 = -\left( {\dot a \over a} \right)^2 [2 + 3\delta + 2 \epsilon + (\delta+\epsilon)(\delta+2\epsilon)]
		- {\dot a \over a}\dot \delta \nonumber \\
	   & = -{\ddot z \over z}
\end{align}
\mynbullet{and}
\myequation{ z \equiv a \left( {\dot a \over a} \right)^{-1} \dot \phi_0\,. }
For any background field, Eqn.~(\ref{eqn:inflationosc}) can be solved to yield the
evolution of the scalar field fluctuation.

\subsection{Slow Roll Limit}

In the slow roll limit $\epsilon, \delta \ll 1$ and the calculation simplifies
dramatically.   The slow roll parameters are usually written in terms of
the derivatives of the potential 
\begin{align}
\epsilon & = {\frac{3}{2}\dot \phi_0^2/ a^2 V  \over 1 + {1 \over 2} \dot \phi_0^2 / a^2 V}
\approx {1 \over 16 \pi G} \left( { V' \over V} \right)^{2} \,,\nonumber\\
\delta & = {\ddot \phi_0 \over \dot \phi_0} \left( {\dot a \over a } \right)^{-1} -1 
\approx  \epsilon - { 1\over 8\pi G }{V''\over V}\,.
\end{align}
The slow roll limit requires a very flat potential.

In the slow roll limit, the
perturbation equation is simply
\myequation{
\ddot u + [k^2  -2\left({\dot a \over a} \right)^2] u = 0\,.}
With the  conformal time measured from the end of inflation
\begin{align}
\tilde \eta &= \eta - \eta_{\rm end}\nonumber\,,\\
\tilde \eta &= \int_{a_{\rm end}}^a {da \over H a^2} \approx -{1 \over a H} \,,
\end{align}
where $a H = \dot a /a$, the equation becomes
\myequation{
\ddot u + [k^2  -{2 \over \tilde \eta^2}] u = 0\,. }
This equation has the exact solution
\myequation{ u = A ( k + {i \over \tilde \eta} ) e^{- i k \tilde\eta} \,,}
where $A$ is a constant.
\mybullet{For $|k\tilde \eta | \gg 1$ (early times, inside Hubble length) the field
 behaves as free
oscillator}
\myequation{ \lim_{|k \tilde \eta| \rightarrow \infty} u= A k e^{- i k \tilde \eta }\,.}
\mybullet{For $|k \tilde \eta| \ll 1$ (late times, $\gg$ Hubble length), the  fluctuation freezes in}
\begin{align*}
 \lim_{|k \tilde \eta| \rightarrow 0 } &u =  {i \over \tilde \eta} A =  {i H aA} \,,\\
& \phi_1 = i H A \,,\\
& \zeta  = -  i H A  \left( {\dot a \over a} \right) {1 \over \dot \phi_0}\,,
\end{align*}
\mybullet{which in the slow-roll approximation can be simplified through}
\begin{align}
\left( {\dot a \over a} \right)^2 {1 \over \dot \phi_0^2} = 
{8 \pi G a^2 V \over 3} {3 \over 2 a^2 V \epsilon}
= {4 \pi G \over \epsilon} = {4 \pi \over m_{\rm pl}^2 \epsilon}\,,
\end{align}
\mybullet{yielding the comoving  curvature power spectrum}
\myequation{
\Delta_\zeta^2 \equiv {k^3 |\zeta|^2 \over 2\pi^2} ={2 k^3 \over \pi} {H^2 \over \epsilon
m_{\rm pl}^2} A^2 \,.}
All that remains is to set the normalization constant $A$ through quantum fluctuations
of the free oscillator.

\subsection{Quantum Fluctuations}

Inside the Hubble length, the classical equation of motion for $u$ is
the simple harmonic oscillator equation
\myequation{ \ddot u + k^2  u =0\,,}
\mybullet{which can be quantized as}
\myequation{ \hat u = u(k,\eta) \hat a + u^*(k,\eta) \hat a^\dagger\,, }
and normalized to zero point fluctuations in the Minkowski vacuum
 $[\hat u , d\hat u/d\eta] = i$,
\myequation{ u(k,\eta) = {1 \over \sqrt{ 2 k} } e^{- i k \tilde \eta}\,. }
Thus $A = (2 k^3)^{1/2}$ and curvature power spectrum becomes
\myequation{
\Delta_\zeta^2 \equiv {1\over \pi} {H^2 \over \epsilon m_{\rm pl}^2} \,. }

The curvature power spectrum is scale invariant to 
the extent that $H$ is constant during inflation.  Evolution in $H$ 
produces a tilt in the spectrum 
\begin{align}
{d \ln \Delta_\zeta^2 \over d \ln k} &\equiv n_S - 1 \nonumber \\
& = 2 {d \ln H \over d\ln k} - {d \ln \epsilon \over d\ln k}\,,
\end{align}
\mybullet{evaluated at Hubble crossing when the fluctuation freezes}
\begin{align}
{d \ln H \over d\ln k}\big|_{-k\tilde \eta = 1} & = {k \over H}{d H \over d \tilde \eta}\big|_{-k\tilde \eta = 1} 
 {d \tilde \eta \over dk}\big|_{-k\tilde \eta = 1}\\
&= {k \over H} (-a H^2 \epsilon )\big|_{-k\tilde \eta = 1} {1 \over k^2} = -\epsilon\,,
\end{align}
\mynbullet{where $a H = -1/\tilde \eta = k$.}
Finally with
\begin{align}
{d \ln \epsilon \over d\ln k} = - {d \ln \epsilon \over d\ln \tilde \eta} =  - 2 (\delta + \epsilon) {\dot a \over a} \tilde \eta = 2(\delta + \epsilon)\,,
\end{align}
\mybullet{the tilt becomes}
\myequation{n_S = 1- 4\epsilon - 2 \delta}
in the slow-roll approximation.

\subsection{Gravitational Waves}

Any nearly massless degree of freedom will acquire quantum fluctuations during inflation.  The inflaton is only special in that it carries the energy density of the universe.
Other degrees of freedom result in isocurvature perturbations. In particular
consider the gravitational wave degrees of freedom.  Their classical equation of motion
resembles the scalar field equation
\begin{equation}
\yell{\ddot H_T^{(\pm 2)}}  +
2 {\dot a \over a}
\yell{\dot H_T^{(\pm 2)}}  +
k^2\yell{H_T^{(\pm 2)}} = 0 \,,
\label{eqn:Poissontensorfreeb}
\end{equation}
\mybullet{and hence acquires fluctuations in same manner as $\phi$}.  
Setting the normalization with
\myequation{\phi_1 \rightarrow H_T^{(\pm 2)} \sqrt{3 \over 16\pi G}\,,}
\mybullet{the gravitational wave power spectrum in each component (polarization
components   $H_T^{(\pm 2)} = (h_+ \pm i h_{\times})/\sqrt{6}$) is }
\myequation{ \Delta^2_H = {16 \pi G \over 3 \cdot 2 \pi^2 } {H^2 \over 2} = {4 \over 3\pi }
{H^2 \over m_{\rm pl}^2 }\,.}

The gravitational wave fluctuations are scale invariant with 
power $\propto H^2 \propto V \propto E_i^4$ where $E_i$
is the energy scale of inflation.  A detection of gravitational waves from inflation
would yield a measurement of the energy scale of inflation.

Finally the tensor tilt
\begin{align}
{d \ln \Delta_H^2 \over d \ln k} &\equiv n_T  = 2 {d \ln H \over d\ln k} = -2\epsilon\,,
\end{align}
\mybullet{which yields a consistency relation between tensor-scalar ratio and tensor tilt}
\myequation{
{\Delta_H^2 \over \Delta^2_\zeta} =  {4 \over 3} \epsilon = -{2 \over 3} n_{T} \,.}

\section{Dark Matter and Energy}
\label{sec:dark}

\subsection{Degrees of Freedom}

A dark component interacts with ordinary matter only through gravity and hence
its observable properties are completely specified by the degrees of freedom in
its stress-energy tensor.  We have seen that without loss of generality these
can be taken as the elements of the symmetric $3 \times 3$ stress tensor. 
Two of these elements represent the scalar degrees of freedom that
influence the formation of structure.  In the background only one of these
remain due to isotropy  leaving only the pressure
$p_{D}$ of the dark component or equivalently the equation of state
$w_{D} = p_{D}/\rho_{D}$. Let us generalize the concept 
of equations of state to the fluctuations in the stress tensor.  

\subsection{Generalized Equations of State}

It is convenient to separate out the non-adiabatic
stress or entropy contribution
\begin{equation}
p_\gdm \Gamma_\gdm =
\delta p_\gdm - c_{\gdm a}^2 \delta \rho_\gdm \,,
\end{equation}
where the adiabatic sound speed is
\begin{equation}
c_{\gdm a}^2 \equiv {\delta p \over \delta \rho}\Big|_{\Gamma_{D}=0} = {\dot p_\gdm \over \dot \rho_\gdm}\,
         = w_\gdm - {1 \over 3} {\dot w_\gdm \over 1 + w_\gdm}
                \left( {\dot a \over a} \right)^{-1} .
\label{eqn:sound}
\end{equation}
Therefore, $p_\gdm = w_\gdm \rho_\gdm$ does {\it not} imply
$\delta p_\gdm = w_\gdm \delta\rho_\gdm$ and furthermore if $\Gamma_\gdm \ne
0$,
the function $w_\gdm(a)$ does not completely specify the pressure fluctuation.
For the dark energy, $w_\gdm < 0$ and is slowly-varying
compared with the expansion rate $(\dot a /a)$
such that $c_{Da}^2 < 0$  The adiabatic pressure fluctuation produces
accelerated collapse rather than support for the density perturbation.
Therefore a dynamical dark energy component must have substantial
non-adiabatic stress fluctuations $\Gamma_{D}\ne 0$ 
 to be phenomenologically viable.

One cannot simply parameterize the pressure fluctuation by a 
non-adiabatic sound speed $\delta p_{\gdm}/\delta \rho_{\gdm}$ since
this is not a gauge invariant quantity and the non-adiabatic stress 
fluctuation is.  The gauge covariant approach allows us to define the
equation of state covariantly.  

As we have seen the comoving gauge gives the closest general 
relativistic analogue to non-relativistic physics.  The generalization of the
comoving gauge to an individual dark component has the conditions 
\begin{equation}
B = v_{\gdm} \,, \qquad H_{T}=0\,,
\end{equation}
which we will call the rest gauge since $T^{0}_{i}=0$.  
Taking
\begin{equation}
{\delta p \over \delta \rho} \Big|_{\rm rest} = c_{\gdm}^{2}\,,
\end{equation}
we obtain via gauge transformation
\begin{align}
\delta p_{\gdm} &= c_{\gdm}^{2}\delta \rho_{\gdm} - ( c_{\gdm}^{2} \dot\rho_{\gdm} - \dot p_{\gdm})
{v_{\gdm}-B \over k} \nonumber\\
&=  c_{\gdm}^{2}\delta \rho_{\gdm} + 3 (1+w_{\gdm}){\dot a \over a}\rho_{\gdm}
( c_{\gdm}^{2} - c_{\gdm a}^{2} )
{v_{\gdm}-B \over k}  \,,
\end{align}
yielding a manifestly gauge invariant non-adiabatic stress
\begin{align}
p_{D}\Gamma_{D} &= (c_{\gdm}^{2}-c_{\gdm a}^{2}) \left[ \delta \rho_{\gdm} 
+  3 (1+w_{\gdm}){\dot a \over a}\rho_{\gdm}
{v_{\gdm}-B \over k} \right] \,.
\end{align}

The anisotropic stress can also affect the density perturbations.
A familiar example is that of a  fluid, where it represents
viscosity and damps density perturbations.  More generally,
the anisotropic stress component is the amplitude of a 3-tensor that
is linear in the perturbation.
A natural choice for its source
is $k v_\gdm$, the amplitude of the velocity shear tensor $\partial^i v^j_\gdm$.
However it must also be gauge invariant and generated by the
corresponding shear term in the metric fluctuation $H_T$.  Gauge transforming
from a gauge with $H_{T}=0$ yields an invariant source of $kv_{\gdm }-\dot H_{T}$.  
These requirements are satisfied by
\begin{equation} \left({d \over d\eta} + 3{\dot a \over a} \right) w_{\gdm} \pi_{\gdm} =
4 c_{\gdm \vis}^2  (k v_\gdm - \dot H_T)\, .
\label{eqn:pi}
\end{equation}

\subsection{Examples}

Cold dark matter provides a trivial example of a dark component.  Here stress fluctuations
are negligible compared with the rest energy density and so $w_{CDM}=c^{2}_{\gdm}=
c^{2}_{D \vis}=0$.  Scalar field dark energy provides another simple example.  
Here the competition between kinetic and potential energy can drive $w_{\phi} < 0$.
As we have already seen a slowly rolling scalar field has
$c_{\phi}^{2}=1$ by virtue of the absence of scalar field fluctuations in the
rest gauge.  In this case the fluctuations bear only the kinetic contributions and so
the relationship is exact.  

The non-adiabatic sound speed $c_{\gdm}^{2}$ is also  useful 
in characterizing $k$-essence, a scalar field with a non-canonical kinetic term.   
Since in the rest gauge the potential fluctuation vanishes, the
sound speed directly reflects the modification to the kinetic term.
A special case occurs when the kinetic term has the wrong sign. 
Here $w_{\phi} <-1$ and the energy density increases as the universe
expands.  This type of matter has been dubbed phantom dark energy.

The viscosity term is relevant for dark radiation components.   Massless neutrinos
can be approximated by $w_{\nu}=c_{\nu}^{2}=c_{\nu \vis}^{2}=1/3$.
A massive neutrino has an equation of state that 
goes from $w_\hdm = 1/3$ to $0$ as the
neutrinos become non-relativistic.  Fitting to the numerical integration of
the distribution gives
\begin{equation}
w_\hdm  = {1 \over 3} \left[ 1+ (a/a_{\rm nr})^{2 p} \right]^{-1/p},
\end{equation}
with $p=0.872$ and $a_{\rm nr}=6.32 \times 10^{-6} /\Omega_\nu h^2$.
We can model its behavior as $c_{\hdm}^{2}=c_{\hdm \vis}^{2}=w_{\gdm}$.

A summary of the phenomenological parameterization of various particle
candidates are given in the table.
\begin{table}
\begin{center}
\begin{tabular}{cccc}
Type & $w_{\gdm}$ & $c_{\gdm}^{2}$ & $c_{\gdm v}^{2}$ \\
\\
CDM & 0 & 0 & 0 \\
$\Lambda$ & -1 & -- & -- \\
Massless $\nu$ & 1/3 & 1/3 & 1/3 \\
Tight coupled $\gamma$ & 1/3 & 1/3 & 0 \\
Hot/Warm DM & $1/3 \rightarrow 0$ & same  &  same \\
Quintessence & variable & 1 & 0 \\
k-essence & variable & variable & 0 \\
Phantom energy & $< -1$ & 1  & 0 \\
Decaying $\nu$ & $1/3 \rightarrow 0 \rightarrow 1/3$ & same & same \\
Axions & 0 & small & 0 \\
Fuzzy Dark Matter &  0 & scale dependent & 0 \\
\end{tabular}
\end{center}
\end{table}

\subsection{Initial Conditions}

With the equations of motion for the dark components defined, all that remains is
to specify the initial conditions. 
Conservation of the comoving curvature outside the Hubble length allows us to ignore
the microphysics of the intermediate reheating phase between inflation and the
radiation dominated universe.  One can then simply take the initial conditions for structure
formation to be in the radiation dominated universe.  Accounting for the neutrino
anisotropic stress as described in the previous section, the initial conditions for the
total perturbations in the comoving gauge become
\begin{align}
\delta & = A_{\delta}(k\eta)^{2}\zeta\,, \nonumber\\
v & = A_{v} (k\eta) \zeta\,, \nonumber\\
\Pi & = A_{\Pi} (k\eta)^{2} \zeta\,,
\end{align}
where $\delta = \delta \rho/\rho$, and $\eta$ is now understood as the conformal
time elapsed after the inflationary epoch.  The constants are 
\begin{align}
A_{\delta} & = {4\over 9}{1+2f_{\nu}/15 \over 1 +4 f_{\nu}/15} (1-3K/k^{2})\,, \nonumber\\
A_{v} & = -{1 \over 3}{1\over 1+4f_{\nu}/15} \,,\nonumber\\
A_{\Pi} & = -{4 \over 15}{1 \over 1+4f_{\nu}/15} \,,
\end{align}
and
\begin{align}
f_{\nu}   &= {\rho_{\nu} \over \rho_{\gamma} + \rho_{\nu}}\,.
\end{align}
The dark component initial conditions can then be determined by detailed balance
\begin{align}
\delta_{J} & = A_{\delta J}(k\eta)^{2}\zeta \nonumber\\
v_{J}-v & = A_{\Delta v J} (k\eta)^{3} \zeta\,, \nonumber\\
\Pi_{J} & = A_{\Pi J} (k\eta)^{2} \zeta
\end{align}
with
\begin{align}
      A_{\Pi J} & = {c_{J\vis}^{2} \over w_{J}} A_{\Pi} \,, \nonumber\\
      A_{\Delta v J } & = {
      {2 \over 15} [2 - 3(w_{J}-c_{J}^{2})]
      (f_{\nu}- {4 \over 1+w_{J}} c_{J\vis}^{2})
      -
      [2(c_{J}^{2}-{1 \over 3}) + (w_{J}-c_{J}^{2})](1+{2\over 15} f_{\nu})
      \over 
      (4 -3 c_{J}^{2} ) [2  - 3(w_{J}-c^{2}_{J})] + 9 (c_{J}^{2} - c_{Ja}^{2}) c_{J}^{2}
      }
      (1-3{K\over k^{2}})A_{v} \,,\nonumber\\
     A_{\delta J} &= {3 \over 4}( 1 + w_{J} ) { A_{\delta}-12(c_{J}^{2}-c_{Ja}^{2}) A_{\Delta v J} 
      \over 1- 3(w_{J} - c_{J}^{2})/2 }\,.              \end{align}
These initial conditions apply to all particle species $J$, including the photons and
neutrinos,
save that for the baryons 
 rapid Thomson scattering with the photons sets $v_{b}=v_{\gamma}$.
In fact for most numerical purposes one can simply set $v_{J}-v=0$ in the initial
conditions and let the
velocity differences arise dynamically in the integration. 
The initial conditions in an arbitrary gauge can be established from these
relations and the gauge transformation properties of the perturbations.

\section{Cosmic Microwave Background}
\label{sec:cmb}

\subsection{Boltzmann Equation}

The gauge covariant formalism is useful for CMB studies as well.  The interpretation
of CMB anisotropy formation is simplest in the Newtonian gauge where the manifestations
of gravitational redshift and infall correspond to Newtonian intuition.  The numerical
solution of the perturbation equations is best handled under the comoving or synchronous
gauge where the fundamental perturbation variables are stable.  The link to 
inflationary initial conditions is best seen in the comoving gauge.   The gauge covariant
approach allows one to calculate in one gauge and interpret in or relate to another.

Most of the physics of CMB perturbation evolution is contained in the general discussion of
\S \ref{sec:formalism}.  However
the cosmic microwave background differs from the dark components in that it undergoes
interactions with the baryons that can exchange energy and momentum.   
Thus the conservation
law for its stress energy tensor must be supplemented with an interaction term. 
CMB observable properties are also not the energy density and velocity perturbations
but the higher order angular distribution of its temperature and polarization.

\mybullet{Generally the CMB is described by the phase space distribution function photons
in each of the two polarization states
 $f_a({\bf x},{\bf q},\eta)$, where ${\bf x}$ is the comoving position 
and ${\bf q}$ is the photon momentum.}
\mybullet{The evolution of the distribution function under gravity and collisions is
governed by the Boltzmann equation}
\myequation{
{d \over d\eta} f _{a,b}= C[f_a,f_b] \,,}
where $C$ denotes the collision term.

In absence of collisions, the 
Boltzmann equation becomes the Liouville equation.  Rewriting the
variables in terms of the photon propagation direction $\bn$,
\begin{align}
{d \over d\eta} f_a({\bf x},\bn,q,\eta) 
&=
\dot f_a + n^{i} \nabla_{i} f_a+ \dot q {\partial \over \partial q} f_a= 0\,.
\end{align}
The last term represents the gravitational redshifting (or Sachs-Wolfe effect) 
of the photons under the  
metric and is given by the geodesic equation as 
\begin{align}
{\dot q \over q} & =
- {\dot a \over a} - {1 \over 2} n^i n^j\dot   H_{T ij}  -  \dot H_L  + n^i \dot B_i  - n^{i} \nabla_{i} A\,.
\end{align}
We have in fact already derived these gravitational effects in the general covariant perturbation
formalism.  To establish this fact note that the stress energy tensor of the photons
is the integral of the photon distribution function over momentum states
\myequation{ T^{\mu \nu} =  \int {d^3 q \over (2\pi)^3 } {q^\mu q^\nu \over E}(f_a + f_b)\,.}
The Liouville equation then expresses the conservation of the stress-energy tensor.
Given that the CMB distribution is observed to be close to blackbody, it is suffices
to calculate the evolution of these integrated quantities. 
In particular let us define the temperature perturbation 
\myequation{
\Theta(\bx,\bn,\eta) = {1 \over 4} \delta_{\gamma} 
= {1 \over 4\rho_\gamma} \int {q^3 d q \over 2\pi^2} (f_a + f_b) - 1}
\mynbullet{and likewise for the linear polarization states $Q$ and $U$ as the temperature
differences between the polarization states.}
The redshift terms then become the metric terms in the continuity and Navier-Stokes
equations.

\subsection{Eigenmodes}

As in the case of the purely spatial 
perturbations, we decompose the temperature and polarization in
normal modes of the spatial and angular distributions.  For the $k$-th spatial eigenmode
\begin{align}
\Theta(\bx,\bn,\eta)      &=  \sum_{\ell m} \Theta_\ell^{(m)} G_\ell^m(\bk,\bx,\bn)\,,\nonumber \\
[Q\pm i U](\bx,\bn,\eta)&=  \sum_{\ell m} [E_\ell^{(m)} \pm i B_\ell^{(m)} ] {}_{\pm 2}G_\ell^m(\bk,\bx,\bn)\,.
\end{align}

\mybullet{They are generated 
by recursion (${}_{0}G=G$)}
\begin{equation}
n^i (\Spin{G}{s}{\ell}{m})_{|i}
        = {q \over 2\ell +1} \left[
          {\Spin{\kappa}{s}{\ell}{m}}  (\Spin{G}{s}{\ell-1}{m})
        - {\Spin{\kappa}{s}{\ell+1}{m}} (\Spin{G}{s}{\ell+1}{m})
        \right]
        - i{q m s \over \ell(\ell+1)}\ \Spin{G}{s}{\ell}{m} \, ,
\label{eqn:recursion}
\end{equation}
where
\begin{align}
q^{2} &= k^{2} + (|m|+1)K \,,\nonumber \\
\Spin{\kappa}{s}{\ell}{m} &= \sqrt{ \left[
{(\ell^2-m^2)(\ell^2-s^2)\over\ell^2}\right]
\left[1 - {\ell^2\over q^2} K \right]}.
\end{align}
The lowest  order modes begin the recursion and are related to the
spatial harmonics as 
\begin{align}
\Gm{0}{j}{m} &= n^{i_1}\ldots n^{i_{|m|}}
        Q_{i_1\ldots i_{|m|}}^{(m)} \,,
        \nonumber \\
\Gm{\pm 2}{2}{m} & \propto  (\hat{m}_1 \pm i \hat{m}_2)^{i_1}
        (\hat{m}_1 \pm i \hat{m}_2)^{i_2}
Q_{i_1 i_2}^{(m)} \,,
\label{eqn:Gl0prime}
\end{align}
where $\hat{\bf m}_{1}$, $\hat{\bf m}_{2}$ span the plane perpendicular to $\hat{\bf n}$. The
normalization is set so that 
\begin{equation}
\Spin{G}{s}{\ell}{m}(0,0,\bn) = (-i)^{\ell} \sqrt{4\pi \over 2\ell +1} \Spin{Y}{s}{\ell}{m} (\bn)\,.
\end{equation}
Here the spin spherical harmonics $\Spin{Y}{s}{\ell}{m}$ are the eigenfunctions
of the 2D Laplace operator on a rank $s$ tensor.   They are given explicitly by rotation
matrices as
\myequation{ {}_s Y_\ell^m(\theta,\phi) = \sqrt{2\ell+1 \over 4\pi} {\cal D}_{-m s}^\ell(\phi,\theta,0)\,.}

The meaning of these modes becomes clear in a spatially flat cosmology.
Here the modes are simply the direct product of plane waves
and spin-spherical harmonics
\begin{align}
G_\ell^m(\bk,\bx,\bn) &\equiv (-i)^\ell \sqrt{ 4\pi \over 2\ell+1} Y_\ell^m(\bn) \exp(i \bk \cdot \bx) 
\nonumber\,, \\
{}_{\pm 2} G_\ell^m(\bk,\bx,\bn) &\equiv (-i)^\ell \sqrt{ 4\pi \over 2\ell+1} {}_{\pm 2} 
Y_\ell^m(\bn) \exp(i \bk \cdot \bx)\,.
\end{align}
\mybullet{The main content of the Liouville equation is purely geometrical and describes the projection
of inhomogeneities into anisotropies.}  Photon propagation takes gradients in the spatial
distribution and converts them to anisotropy as
\myequation{ \bn \cdot \nabla e^{i \bk \cdot \bx} = i \bn \cdot \bk e^{i \bk \cdot \bx} =
i \sqrt{4 \pi \over 3} k Y_1^0(\bn) e^{i \bk \cdot \bx}\,.}
\mybullet{This dipole term adds to angular dependence through the addition of angular momentum}
\myequation{
\sqrt{4 \pi \over 3} Y_1^0 Y_\ell^m = {\kappa_\ell^m \over \sqrt{(2\ell+1)(2\ell-1)}} Y_{\ell-1}^m
	+ {\kappa_{\ell+1}^m \over \sqrt{(2\ell+1)(2\ell+3)}} Y_{\ell+1}^m \,,}
	\mynbullet{where $\kappa_\ell^m = \Spin{\kappa}{s}{\ell}{m} = \sqrt{\ell^2 - m^2}$ is related to the Clebsch-Gordon 
	coefficients.}

\subsection{Collision Term}

The dominant collision process for CMB photons is Thomson scattering off of
free electrons which has the differential cross section
\myequation{
{d \sigma \over d\Omega} = {3 \over 8\pi} |\hat {\bf E}' \cdot \hat{\bf E}|^2\sigma_T
\,,
\label{eqn:Thomson}
}
\mynbullet{where $\hat {\bf E}'$ and $\hat {\bf E}$ denote the incoming and outgoing
directions of the electric field or polarization vector.}

To evaluate the collision term we begin in 
 the electron rest frame and in a coordinate system fixed by the scattering plane,
spanned by incoming and outgoing directional vectors $-\bn' \cdot \bn = \cos \beta$, where
$\beta$ is the scattering angle.
Denoting $\Theta_\parallel$ as the in-plane polarization temperature fluctuation and $\Theta_\perp$ as the perpendicular polarization state, we obtain the geometrical 
content of the transfer equation
\begin{align}
\Theta_\parallel & \propto \cos^2 \beta \, \Theta_\parallel' , \qquad
\Theta_\perp        \propto \Theta_\perp'\,,
\end{align}
where the proportionality reflects the scattering rate 
\begin{equation}
\dot \tau = n_{e}\sigma_{T}a \,. 
\end{equation}

To calculate the Stokes parameters in this basis, we also need to calculate
the polarization states with axes rotated by $45^{\circ}$
\begin{align}
\hat{\bf E}_1 = {1 \over \sqrt{2}} (\hat{\bf E}_\parallel + \hat{\bf E}_\perp)\,, \qquad
\hat{\bf E}_2 = {1 \over \sqrt{2}} (\hat{\bf E}_\parallel -  \hat{\bf E}_\perp)\,,
\end{align}
yielding the transfer
\begin{align}
\Theta_1 & \propto |\hat{\bf E}_1 \cdot \hat{\bf E}_1|^2 \Theta_1' +
			      |\hat{\bf E}_1 \cdot \hat{\bf E}_2|^2 \Theta_2'  \nonumber\\
		& \propto {1 \over 4}(\cos\beta +1)^2 \Theta_1' +
		       {1 \over 4}(\cos\beta - 1)^2 \Theta_2' \nonumber\\
\Theta_2 & \propto |\hat{\bf E}_2 \cdot \hat{\bf E}_2|^2 \Theta_2' +
			      |\hat{\bf E}_2 \cdot \hat{\bf E}_1|^2 \Theta_1'  \nonumber \\
		& \propto {1 \over 4}(\cos\beta +1)^2 \Theta_2' +
		       {1 \over 4}(\cos\beta - 1)^2 \Theta_1'	\,.	
\end{align}
Now the transfer properties of the Stokes parameters 
\begin{align}
\Theta \equiv{1 \over 2} ( \Theta_\parallel + \Theta_\perp), \quad 
Q \equiv {1 \over 2}(\Theta_\parallel - \Theta_\perp), \quad
U \equiv {1 \over 2}(\Theta_1 - \Theta_2)
\end{align}
arranged in a vector 
 ${\bf T} \equiv$ ($\Theta$, $Q + i U$, $Q - i U$) becomes
\myequation{
{\bf T} \propto {\bf S}(\beta) {\bf T}' \,,}
\begin{align}
\nonumber
{\bf S}(\beta) =
{3 \over 4}\left(
\begin{array}{ccc}
\cos^2\beta +1  \quad & -{1 \over 2}\sin^2\beta  \quad
                       & -{1 \over 2}\sin^2\beta \vertsp\\
-{1 \over 2}\sin^2  \beta
                 \quad & {1 \over 2}(\cos\beta + 1)^2 \quad
                       & {1 \over 2}(\cos\beta - 1)^2 \vertsp\\
-{1 \over 2}\sin^2 \beta
                 \quad & {1 \over 2}(\cos\beta - 1)^2 \quad
                       & {1 \over 2}(\cos\beta + 1)^2 \vertsp\\
\end{array}
\right)\,,
\end{align}
\mynbullet{where the normalization factor is set by photon conservation in the 
scattering.}

Finally convert the the polarization quantities referenced to the scattering basis
to a fixed basis on the sky by noting that under a rotation 
${\bf T}' = {\bf R}(\psi) {\bf T}$ where 
\begin{align}
{\bf R}(\psi) = \left(
\begin{array}{ccc}
1 & 0 & 0 \vertsp\\
0 & e^{-2i\psi} & 0 \vertsp \\
0 & 0 & e^{2i\psi} \vertsp 
\end{array} \right)\,,
\end{align}
\mynbullet{giving the scattering matrix}
\small\begin{align}
&{\bf R}(-\gamma){\bf S}(\beta){\bf R}(\alpha)  =\nonumber\\
&\quad
{1 \over 2}{\sqrt{4\pi \over 5}}\left(
\begin{array}{ccc}
Y_2^0(\beta,\alpha)  + 2 \sqrt{5} Y_0^0(\beta,\alpha) \quad
        & -\sqrt{3 \over 2} Y_2^{-2}(\beta,\alpha) \quad
        & -\sqrt{3 \over 2} Y_2^2(\beta,\alpha) \vertsp\\
-\sqrt{6} \Spy{2}{2}{0}(\beta,\alpha)e^{2i\gamma} \quad
        & 3 \Spy{2}{2}{-2}(\beta,\alpha)e^{2i\gamma} \quad
        & 3 \Spy{2}{2}{2}(\beta,\alpha)e^{2i\gamma} \vertsp\\
-\sqrt{6} \Spy{-2}{2}{0}(\beta,\alpha)e^{-2i\gamma} \quad
        & 3 \Spy{-2}{2}{-2}(\beta,\alpha)e^{-2i\gamma} \quad
        & 3 \Spy{-2}{2}{2}(\beta,\alpha)e^{-2i\gamma} \vertsp
\end{array}
\right) \,,
\end{align}
where $\alpha, \gamma$ are the angles required to rotate into and out of
the scattering frame.  

Finally, by employing the addition theorem for spin spherical harmonics
\begin{align}
\sum_m \Spy{s_1}{\ell}{m*} (\bn') \,\,
        \Spy{s_2}{\ell}{m}(\bn)  
= (-1)^{s_1-s_2}\sqrt{2\ell+1 \over 4\pi}
 \,\, \Spy{s_2}{\ell}{-s_1} (\beta,\alpha)
e^{i s_2 \gamma} 
\end{align}
the scattering in the electron rest frame into the Stokes states becomes
\begin{align}
{C}_{\rm in}
[{{\bf T}}]&=  \dot\tau \int {d\bn' \over 4\pi}
 {\bf R}(-\gamma){\bf S}(\beta){\bf R}(\alpha) {\bf T}(\bn')\nonumber \\
& =  \dot\tau \int {d\bn' \over 4\pi} (\Theta',0,0) +
        {1 \over 10}\dot\tau \int d\bn'
        \sum_{m=-2}^2 {\bf P}^{(m)}(\bn,\bn') {\bf T}(\bn')\, ,
\label{eqn:fullcollision}
\end{align}
\mynbullet{where the quadrupole coupling term is}
\small\begin{align}
{\bf P}^{(m)}(\bn,\bn')=
\left(
\begin{array}{ccc}
Y_2^{m*}(\bn')\, Y_2^m(\bn)   &
          - \sqrt{3 \over 2} \Spy{2}{2}{m*}(\bn')\, Y_2^m(\bn)
         & - \sqrt{3 \over 2}
          \Spy{-2}{2}{m*}(\bn')\, Y_2^m(\bn)
        \vertsp\\
- \sqrt{6} Y_2^{m*}(\bn') \Spy{2}{2}{m}(\bn)  &
3 \Spy{2}{2}{m*}(\bn')\Spy{2}{2}{m}(\bn)  &
          3 \Spy{-2}{2}{m*}(\bn')\Spy{2}{2}{m}(\bn) 
                \vertsp\\
- \sqrt{6} Y_2^{m*}(\bn') \Spy{-2}{2}{m}(\bn)  &
          3 \Spy{2}{2}{m*}(\bn') \Spy{-2}{2}{m}(\bn) &
          3 \Spy{-2}{2}{m*}(\bn') \Spy{-2}{2}{m}(\bn)
                \vertsp\\
\end{array}
\right) \,.
\end{align}
\mybullet{The full scattering matrix involves difference of scattering into and out of state}
\begin{align}
{C}[{{\bf T}}] = {C}_{\rm in}[{{\bf T}}] - {C}_{\rm out}[{{\bf T}}]\,.
\end{align}
\mybullet{In the electron rest frame}
\begin{align}
{C}[{{\bf T}}] = \dot \tau  \int {d\bn' \over 4\pi} (\Theta',0,0)  -\dot\tau {\bf T} + C_{P}[{\bf T}]
\end{align}
\mynbullet{which describes isotropization in the rest frame.  All moments have $e^{-\tau}$
suppression except for isotropic temperature $\Theta_0$. Here $C_{P}$ is the  $P$ or
$\ell=2$ term of Eqn.~(\ref{eqn:fullcollision}).  

Transformation into the background frame simply induces a dipole term}
\begin{align}
{C}[{{\bf T}}] = \dot \tau  \left( \bn \cdot {\bf v}_b + \int {d\bn' \over 4\pi} \Theta',0,0\right)  -\dot\tau {\bf T} + C_P[{\bf T}]\,,
\end{align}
yielding the final form of the collision term.

\subsection{Temperature-Polarization Hierarchy}

\mybullet{The Boltzmann equation in normal modes then becomes}
\begin{align}
\dot \Theta_\ell^{(m)} & = q \left[ { \kappa_\ell^m \over 2\ell+1} \Theta_{\ell-1}^{(m)} - 
					      { \kappa_{\ell+1}^m \over 2\ell+3} \Theta_{\ell+1}^{(m)} \right] 
						- \dot\tau \Theta_\ell^{(m)} +  S_\ell^{(m)}\,, \nonumber\\
\dot E_\ell^{(m)} &= k \left[ { {}_2 \kappa_\ell^m \over 2\ell-1 } E_{\ell-1}^{(m)} 
		- {2 m \over \ell(\ell+1)} B_\ell^{(m)} - { {}_2 \kappa_{\ell+1}^m \over 2\ell+3 } \right]
		-\dot \tau E_\ell^{(m)} + {\cal E}_\ell^{(m)}\,,\nonumber \\
\dot B_\ell^{(m)} & = k \left[ { {}_2 \kappa_\ell^m \over 2\ell-1 } B_{\ell-1}^{(m)} 
		+ {2 m \over \ell(\ell+1)} B_\ell^{(m)} - { {}_2 \kappa_{\ell+1}^m \over 2\ell+3 } \right]
		-\dot \tau E_\ell^{(m)} + {\cal B}_\ell^{(m)}\,,
\end{align}
where the gravitational and scattering sources are
\begin{align}
S_{\ell}^{(m)} & =
\left(
\begin{array}{lll}
\dot\tau \Theta_0^{(0)} -  \dot H_L^{(0)}  &
\dot\tau v_b^{(0)} + \dot B^{(0)} &
\dot\tau P^{(0)}  -{2 \over 3} \sqrt{1 -3K/k^{2}} \dot H_T^{(0)}
         \vertsp\\
                                                0 &
\dot\tau v_b^{(\pm 1)} + \dot B^{(\pm 1)}&
\dot\tau P^{(\pm 1)} -{\sqrt{3} \over 3} \sqrt{1-2K/k^{2}} \dot H_T^{(\pm 1)}
        \vertsp\\
                                                0 &
                                                0 &
\dot\tau P^{(\pm 2)} - \dot H_T^{(\pm 2)}      \vertsp 
\end{array}\right)\nonumber\,,\\
{\cal E}_\ell^{(m)} &= -\dot \tau \sqrt{6} P^{(m)} \delta_{\ell,2} \,,\nonumber\\
{\cal B}_\ell^{(m)} &= 0 \,,
\end{align}
\mynbullet{with}
\myequation{P^{(m)} \equiv {1 \over 10} (\Theta_2^{(m)} -\sqrt{6} E_2^{(m)} )\,.}
The physical content of the coupling hierarchy is that an inhomogeneity in the temperature
 or polarization distribution will eventually become a high multipole order anisotropy by
 ``free streaming" or simple projection.

\subsection{Integral Solution}

\mybullet{Since the hierarchy equations simply represents geometric projection,
their implicit solution can be written as the projection of the gravitational and scattering
sources at a distance.   This operation proceeds by writing the normal modes themselves
in spherical coordinates,
\begin{align}
G_{\ell_s}^m &= \sum_\ell (-i)^\ell \sqrt{ 4\pi (2\ell+1)} \alpha_{\ell_s \ell}^{(m)}(k,D) 
						Y_\ell^m(\bn) \,,\nonumber\\
{}_{\pm 2} G_{\ell_s}^m &= \sum_\ell (-i)^\ell \sqrt{ 4\pi (2\ell+1)} [\epsilon_{\ell_s \ell}^{(m)}
\pm \beta_{\ell_s \ell}^{(m)}](k,D) 
						\Spin{Y}{\pm 2}{\ell}{m}(\bn) \,,
\end{align}						
where $D= \eta_{0}-\eta$.  Summing over the sources
\begin{align}
{\Theta_\ell^{(m)}(k,\eta_0) \over 2\ell +1 } &= \int_0^{\eta_0} d\eta e^{-\tau}
\sum_{\ell_s} S_{\ell_s}^{(m)} \alpha_{\ell_s \ell}^{(m)}(k,D) \,,\nonumber\\
{E_\ell^{(m)}(k,\eta_0) \over 2\ell+1} &= \int_0^{\eta_0} d\eta e^{-\tau} 
\sum_{\ell_{s}}{\cal E}_{\ell_s}^{(m)}
				\epsilon_{\ell_s \ell}^{(m)}(k,D)  \,,\nonumber \\
{B_\ell^{(m)}(k,\eta_0) \over 2\ell+1} &= \int_0^{\eta_0} d\eta e^{-\tau} 
\sum_{\ell_{s}}
{\cal E}_{\ell_s}^{(m)}
				\beta_{\ell_s \ell}^{(m)}(k,D)  \,.
\end{align}
Note that the polarization has only an $\ell_{s}=2$ source.

In a flat cosmology, the radial projection kernels are related to spherical Bessel functions
\myequation{e^{i \bk \cdot \bx} = \sum_{\ell} (-i)^\ell \sqrt{4\pi (2\ell+1) } j_\ell(k D) Y_\ell^0(\bn)}
by the recoupling of the ``spin'' angular dependence of the source $\ell_{s}$ to the
``orbital''  angular dependence of the plane waves.  For example
\begin{align} 
 \alpha_{0\ell}^{(0)}(k,D) &\equiv j_\ell(kD) \,,\nonumber\\
   \alpha_{1\ell}^{(0)}(k,D) &\equiv j_\ell' (kD)\,,\nonumber\\
 \epsilon_{2\ell}^{(0)}(k,D) &= \sqrt{ {3\over 8}{(\ell+2)! \over (\ell-2)!}} {j_\ell(kD) \over (kD)^2} \,,\nonumber\\
\beta_{2\ell}^{(0)}(k,D) &= 0\,.
\end{align}
In a curved geometry, the radial projection kernels are related to the ultraspherical Bessel
functions by the same coupling of angular momenta. 

\subsection{Power Spectra}

The two point statistics of the temperature and polarization fields are described
by their power spectra
\begin{align}
C_\ell^{X X'}} &= {2 \over \pi } \int{ dk \over k} \sum_m {k^3 \langle X_\ell^{(m)*} X_\ell^{'(m)}\rangle
\over (2\ell+1)^2} \,,
\end{align}
where $X,X' \in \Theta, E, B$.

\section{Epilogue}

How to conclude {\it Lecture notes} in which no conclusions have been drawn?  I leave
you instead with a thought:
\begin{quote}
What goes on being hateful about analysis is that it implies that the analyzed
is a completed set.   The reason why completion goes on being hateful is that
it implies everything can be a completed set.
\vskip 0.1truecm
\centerline{--Chuang-tzu, 23}
\end{quote}

\section{Acknowledgments}

I thank the organizers and participants of the Trieste school, my collaborators over the years 
(especially D. Eisenstein, N. Sugiyama and M. White in this context), and 
the long-suffering students of AST448 at U. Chicago.


\end{document}